# The Need to Reconcile Concepts that Characterize Systems Facing Threats

Stephanie Galaitsi[1], Jeffrey, M. Keisler[2], Benjamin D. Trump[1], and Igor Linkov[1*]

[1] US Army Corps of Engineers.

[2] University of Massachusetts, Boston, Boston, MA, USA.

* Address Correspondence to Igor Linkov; Igor.Linkov@usace.army.mil.

Desirable system performance in the face of threats has been characterized by various management concepts. Through semi-structured interviews with editors of journals in the fields of emergency response and systems management, a literature review, and professional judgment, we identified nine related and often interchangeably-used system performance concepts: adaptability, agility, reliability, resilience, resistance, robustness, safety, security, and sustainability. A better understanding of these concepts will allow system planners to pursue management strategies best suited to their unique system dynamics and specific objectives of good performance. We analyze expert responses and review the linguistic definitions and mathematical framing of these concepts to understand their applications. We find a lack of consensus on their usage between interview subjects, but by using the mathematical framing to enrich the linguistic definitions, we formulate comparative visualizations and propose distinct definitions for the nine concepts. We present a conceptual framing to relate the concepts for management purposes.

**KEY WORDS:**

## 1. INTRODUCTION

Today's natural and human-built world comprises complex and interconnected systems. Such systems can encounter various forms of threats, challenges, or disruptions, and a system's subsequent performance can be characterized by many concepts, including adaptability, agility, reliability, resilience, resistance, robustness, safety, security, and sustainability. The universal objective of "good" performance has produced ambiguity over the various means for achieving it. The Federal Emergency Management Agency (FEMA) supports community *resilience* in the face of disasters, and the Center for Disease Control studies bacterial *resistance* to antibiotics. Organizations benefit from *agility*, and the US military-industrial complex seeks to foster *robustness*, *security*, and *resilience*. The US Federal Motor Carrier *Safety* Administration prevents commercial motor vehicle-related fatalities and injuries. National parks should be *sustainable*, while critical infrastructure and cyberspace demand *security*, the City of Boston is developing its *resilience* and *adaptability*, and electricity grids value *reliability*.

There is no consensus across fields regarding relationships between these concepts. A review (SI1) reveals that while many papers compare two or three concepts, there is no comprehensive evaluation of their distinct roles in relation to the overall system performance.

The concepts themselves are not always consistently defined or evaluated. Taking resilience as one example, several review papers find multiple and often competing definitions of resilience (e.g., Wied, Oehmen, & Welo, 2020; Hosseini, Barker, & Ramirez-Marquez, 2016; Southwick, Bonanno, Masten, Panter-Brick, & Yehuda, 2014; Linkov & Trump, 2019). Ganin et al. (2016) define resilience as the ability to recover and retain critical system functionality in response to a wide range of threats and Yonson and Noy (2019) quantify resilience as the ability to minimize losses following a disaster and also by the ability to reconstruct and recover quickly. Gillespie-Marthaler, Nelson, Baroud, and Abkowitz (2019) define sustainable resilience as maintaining desired system performance while considering instrasystem and intergenerational distribution of capital or consequent vulnerabilities. Yu and Baroud (2019) measure community resilience by the rate of recovery, which can be predicted according Bayesian analyses of past recoveries and Rand, Kurth, Fleming, and Linkov (2020) look at resilience quantification as multicriteria problem. Resilience may be defined using other system concepts: Donahue et al. (2016) found characterizations of resilience using resistance and robustness and sustainability. Specking et al. (2019) relate resilience to "ilities" like reliability, availability, and maintainability that help the system achieve performance through accomplishing its designed tasks. Creaco, Franchini, and Todini (2016) found resilience to be part of an indirect measure of reliability. Marchese et al. (2017) identified three frameworks relating resilience and sustainability: those that claimed resilience was a component of sustainability, those claiming sustainability was a component of resilience, and those claiming the concepts are unrelated.

The frequent and inconsistent usage of these terms risks mischaracterizing system management objectives. For managers seeking to improve a system's ability to continue functioning through disruptions, the system performance expectations and available strategies for supporting it should be clearly differentiated. As human-managed systems increasingly support various societal activities, funding allocations should reflect the strengths, weaknesses, and opportunities embodied in various system characteristics. Such funding appropriation becomes difficult when the interpretations of these concepts vary between disciplines. To avoid redundant, counterproductive, or unnecessarily costly efforts, concepts for system responses to threats, challenges, or disruptions require clear distinctions.

To examine the magnitude of disconnect in the field, we conducted semi-structured interviews with editors from the top journals of fields related to risk, disasters, and safety and found that the interpretation of the nine concepts differs widely across the very people who define these fields. We follow Boholm (2019) by conducting a linguistic analysis, in our case to examine the terms' definitions and etymologies to identify areas of clear differentiation as well as overlaps. Finally we identified examples of first principle mathematical framing of these concepts to further substantiate differentiations. We propose a conceptual framing that synthesizes the differences and relationships that we identified, which can support communication between strategic system planners of different fields and facilitate transferring successful management strategies. To our knowledge, such an overarching analysis has not been previously provided.

## 2. SYNTHESIZING EXPERT VIEWS ON THEMATIC CONCEPTS OF SYSTEM

We conducted semi-structured interviews with eleven experts from systems management fields. To identify journals, we used the Web of Science to search multiple terms related to systems management, risk, disasters, safety, and other relevant concepts. Ten editors were contacted, and through soliciting additional names from the editors who responded, we created a purposeful sample that yielded 11 interview responses, including nine editors of academic journals (impact factor > 1) or their designees and two NSF program directors.

For the interviews, we hypothesized that each concept implies certain characteristics about both the disruption encountered and the system's reaction to it. We asked the interview subjects to classify each term according to three dichotomies:

1. Chronic vs. acute threats or disruptions

      2. Focused vs. broad system response
      3. Internal vs. external system response

    Chronic threats are ordinary, repetitive, and expected over the planning horizon, whereas acute threats are comparably extreme, rare, and perhaps unexpected events that may or may not occur during the planning horizon. The boundary between chronic and acute threats can change over time, such as the case of flooding risks under climate change. Next, a focused system response is capable of absorbing a specific type of threat, while a broad response addresses many potential threats, including new or unknown threats. Finally, we asked whether the system's threats are absorbed externally by an ancillary protection (e.g. a soldier's shield), or internally, by an integrated component of the system (e.g., a human immune system). The informal discussions also solicited additional concepts for our list.

    Table I includes a summary of integrated interview responses. Detailed results and explanations of the analysis are presented in SI2. Cells marked with X show the experts' most common response (colored green) to the question of which aspect of a dichotomy best characterizes each term. The color saturation difference between red and green indicates the strength of the responses towards one aspect of the dichotomy across the experts. Responses that were tied (no coherency) are shown in yellow.

Table I. Results of Semi-Structured Interviews

|  | Threat Type | | System Response | | | |
|---|---|---|---|---|---|---|
|  | Acute | Chronic | Focused | Broad | External | Internal |
| Adaptability |  | X |  | X |  | X |
| Agility | X |  | X |  |  | X |
| Reliability |  | X | X |  |  | X |
| Resilience | X |  |  | X |  |  |
| Resistance | X |  | X |  | X |  |
| Robustness | X |  | X |  | X |  |
| Safety |  | X | X |  | X |  |
| Security |  | X | X |  | X |  |
| Sustainability |  | X |  |  | X |  |

    The results of the semi-structured interviews show little consensus. Responses showed consistency around two concepts: systems characterized by *sustainability* were viewed almost unanimously to be coping with chronic threats, and *adaptable* responses were mostly considered to be broad defenses. Safety and security averaged the same categorizations in all three dichotomies, as did resistance and robustness, though they all differed within individual interviews. Many experts found the proposed dichotomies irrelevant to differentiate between the concepts, sometimes granting all concepts the same score in a given dichotomy. One expert noted that the dichotomies were so tenuously related to his idea of the concepts that he theorized he might evaluate them differently were he asked a second time. Expert suggestions for additional dichotomies included long-term/short-term and rote vs. creative responses.

    These results indicate that a basis for clear differentiation between the concepts will not be conclusively found within the usages of current scholarship, and reflects a general lack of coherence seen in the literature. The National Science Foundation has dedicated considerable effort to defining these concepts already without a clear consensus. As buzzwords like resilience and sustainability emerge in response to new realities and challenges, some confusion and interpretation should be expected. These terms are active in English vernacular and can be applied to systems analysis for a variety of reasons that may deviate from an accurate reflection of the term's actual meaning. In gerontology, for example, models for *resilience* in aging arose from critiques of the existing model of successful aging (Cosco, Wister, Brayne, & Howse, 2018), and the resilience application introduced therein does not greatly resemble

resilience applications used elsewhere, such as the National Academy of Science definition for resilience to natural disasters (Linkov, I., Klasa, K., Galaitsi, S., & Wister, A., 2020). When a field begins using a term, its usage can reflect hysteresis in the field's existing terminology and the need for innovation rather than a close match with a term's actual meaning in other contexts. Thus, as more fields use more terms, the overarching terminology meanings can become increasingly muddled.

Reconciling the usage of these concepts and their relationships to each other might require some fields to relinquish their current terminological frameworks, which would not be defensible without a firm basis in existing consensus usages. This does not, however, diminish the importance to facilitate communication and to reconcile the concepts in their applications everywhere, but it does emphasize the difficulty of the task. There is a lost opportunity for efficiency and learning, and an increased chance of misunderstanding in communications between scientists, funders, and decision makers. Thus our analysis now turns to other sources of information: linguistic definitions followed by mathematic framings.

# 3. CRITICAL EXAMINATION OF LINGUISTIC CHARACTERISTICS

Our linguistic exploration uses two readily available sources: the online Oxford Lexico Dictionary (lexico.com) and the Online Etymology Dictionary (etymonline.com). For consistency, we use the first noun definition provided for all the terms with the exceptions of robustness and resistance, which use the dictionary's system-specific definitions. Next, we list relevant adjective and verb forms of each term. Then the etymologies consider the roots and original meaning of each term, including splitting the word into its components where applicable. Finally, the last column lists our interpretations of the definitions, grammar usages, and etymologies, which are summarized in the text following Table II and provided in full in SI2.

**Table II.** Linguistic Characteristics

| Noun | Oxford Definition | Adjective | Verb | Etymology | Interpretation |
|---|---|---|---|---|---|
| Adaptability | The quality of being able to adjust to new conditions. | Adaptable | Adapt | **VERB** Latin: *ad* + *aptare* to[make] + to fit | Reaction |
| Agility | Ability to move quickly and easily. | Agile | | **ADJECTIVE** Latin: *agilis* nimble, quick | Action |
| Reliability | The quality of being trustworthy or of performing consistently well. | Reliable | Rely | **VERB** Latin: *re* + *ligare* again + to bind | Overall performance |
| Resilience | The capacity to recover quickly from difficulties; toughness. | Resilient | Resile [rare] | **VERB** Latin: *re* + *salire* again + to leap | Reaction |
| Resistance | The ability not to be affected by something, especially adversely. | Resistant | Resist | **VERB** Latin: *re* + *sistere* again + stand | Reaction |
| Robustness | The ability to withstand or overcome adverse conditions or rigorous testing. | Robust | | **ADJECTIVE** Latin: *robustus* strong and hardy | Reaction |
| Safety | The condition of being protected from or unlikely to cause danger, risk, or injury. | Safe | | **ADJECTIVE** Latin: *salvus* uninjured, in good health, safe | State/ Condition |
| Security | The state of being free from danger or threat. | Secure | Secure | **ADJECTIVE** Latin: *se* + *cura* free from + care | State/ Condition |
| Sustainability | The ability to be maintained at a certain rate or level. | Sustainable | Sustain | **VERB** Latin: *sub* + *tenere* from below + to hold | Overall performance |

The noun definitions contain elements providing mutual exclusivity, with the possible exception of resistance and robustness. For our interpretations, we found that some concepts characterize actions without reference to stimulus (agility), while others reference a change or discontinuity in conditions that necessitate reactions or responses on the part of the system (resistance, robustness, resilience, adaptability). Sustainability and reliability refer to overall performance, and safety and security are defined as a state or condition, which can be synonyms. These categorizations do not correspond to the qualifiers in the definitions (quality, ability, capacity, etc.) except in the case of safety and security.

Five etymologies stem from verbs (adaptability, resilience, resistance, reliability, sustainability), and the remaining four from adjectives (security, safety, agility, and robustness). All exist as adjectives and nouns in modern form, but only five comprise commonly-used verbs. From this, we conclude that it is easier to characterize a concept (verb becoming an adjective) than to describe what it does (adjective becoming verb). SI3 examines the six verbal forms (including resile) of the concepts within sentence structure and found that three (adapt, resile, resist) modify subjects while three (rely, secure, sustain) modify objects. This suggests more direct actionable control over the concepts of adaptation, resilience, and resistance, and less direct control over reliability, security, and sustainability.

According to the Online Etymology Dictionary, all nine words stem from Latin roots, with some further traceable to Proto-Indo-European origins. Thus, the words have existed simultaneously for millennia.  There is the notable repetition of the prefix "re" in resistance, resilience, and reliability, from the same Latin root meaning *again*, which is closely associated with the word *against.* The remaining prefixes include a verb, making adaptability highly action-oriented, contrasted with an adjective modifying a noun (se + cura), and preposition modifying a verb (sub + tenere), both of which denote more steady state. Thus the prefixes suggest actions (adaptability), reactions (reliability, resilience, resistance) and more steady states (security, sustainability).

Finally, the noun forms of the concepts take particular suffixes,–*ability*, –*ance/-ence, -ity* and –*ty.* We note that the suffixes do not directly correspond with the assignments of quality/capacity/ability/state/ condition provided by the definitions. Suffixes affect a noun's meaning. The nouns *sustenance* and *reliance* are markedly different concepts from *sustainability* and *reliability*. The –*ance* suffixes may denote an earned or acquired resource that can be reliably deployed (see intelligence vs. intelligibility); thus *resilience* and *resistance* may denote some permanence in characterization or assurance of action. This is contrasted with –*ability* suffixes that more indicate nebulous future potential or uncertain characterization. The -*ity* and –*ty* suffixes imply a state (see immunity vs. immunization, or indignity vs. indignation), indicating more certainty in characteristic.

The grammar, prefix, and suffix characterizations do not perfectly align with our interpretations of the modern definitions, but there is sufficient overlap (see Table in SI3) to suggest that modern definitions capture the variability between these different linguistic analyses. Thus, we assert that the vernacular division of action, reaction, state/condition, and overall performance is the most informative interpretation of the different concepts.

## 4. MATHEMATICAL FRAMING

The linguistic presentations of the concepts suggest that they can be evaluated and measured for natural phenomena and systems operations. But while heuristic metrics based on checklists and the like abound, definitions from first principles are somewhat rare in the scientific literature. Where such definitions are found, they tend to reside in a discipline for which the concept is essential, often to the exclusion of other disciplines. In identifying and examining such scientific formulations, we found that these mathematical framings substantiate the vernacular for the concepts. We approach this analysis systemically: these quantitative definitions can provide a foundation for differentiating the concepts according to particular system characteristics. Though our searches were not comprehensive, the results show consonance and provide insight to the specific contributions of each concept within a systemic context. Table III presents condensed versions of the definitions and categorical interpretation which relate to general systems.

The definitions of Table III allow us to populate the other columns. Firstly, the dichotomy of acute vs. chronic is not clearly differentiated because multiple acute threats could occur, and for some concepts, like reliability or sustainability, the concepts do not necessarily imply an outside threat so much as inherent internal variation or degradation, for reasons unknown. Therefore, we find that two categories of disruption type existed: 1) external threat with acute or chronic impact, and 2) unspecified threat origin with chronic impact. We next find that the concepts characterize different periods within the course of a disruptive event. Some correspond to the disruption's impact on the system's performance, whereas others characterize the system's response after the worst of the damage has occurred, or both. Next we propose an equation, parsed into a numerator and a denominator. The numerator includes the more unique parameters of each term, for example the numerator of reliability is the ratio of values within thresholds to values outside thresholds. The denominator is expressed either as per event, or over time, with the exception of robustness and security, for which it is not necessary to specify. The denominators that use "event" could certainly be averaged, summed, or otherwise combined to characterize performance over time, but also stand on their own, while the concepts with time as the denominator cannot be adequately characterized by a single disruption event.

**Table III.** Mathematical Framing of Concepts

| Concept | Mathematical Framing | Threat Type | Impact/ Recovery | Proposed Equation Numerator | Denominator |
|---|---|---|---|---|---|
| Adaptability | The probability that a population will be able to achieve a given level of fitness after a disruption, e.g., a population with limited genetic variability will not be able to adapt (even after the initial die-off) as much as one that has more variability (Stewart, Parsons, & Plotkin, 2012). This has been generalized to the ability of biosystems to recover functionality (Ch'ng, 2007). | External, acute or chronic | Both | Population x Probability of (necessary-to-survive) individual change | (Per) Event |
| Agility | The ability to execute rapid whole body movement with change of velocity or direction in response to a stimulus (Sheppard & Young, 2006). | External, acute or chronic | N/A | Response time ÷ Response time needed to avoid penalty | (Per) Event |
| Reliability | The percentage of time that a system function remains within a threshold of acceptable performance. For voltage, some variation is unavoidable but too much variation causes problems. In systems engineering, reliability may be associated with a probability of failure over a given time period under typical operating conditions. Any quantitative measure of reliability depends on both the variability of the system and the tightness of its thresholds. | Unspecified threat origin, chronic | Both | Values within thresholds ÷ values outside thresholds | (Over) Time |
| Resilience | A perfectly resilient rubber ball is one that bounces back to the same height it is dropped from. A softer rubber ball may return to its original shape less quickly, dissipating energy through the process, and subsequently bounce to a lower height with less potential energy. Thus, resilience in the ball preserves energy through quick and full recovery. | External, acute or chronic | Recovery | Percent critical function recovery ÷ recovery time | (Per) Event |
| Resistance | The ratio of electric pressure (Volts) to the amount of current (Amps). Resistance characterizes the system's ability to absorb pressure while minimizing the impact on the system. | External, acute or chronic | Impact | Pressure ÷ damage to system | (Per) Event |
| Robustness | The amount of pressure the system can withstand without any change in functionality. This appears in various types of engineering: "Robustness is the ability of the system to avoid failure modes, even in the presence of realistic noises" (Clausing, 2004) or "the ability of a system to resist change without adapting its initial stable configuration" (Wieland & Wallenburg, 2012). | External, acute or chronic | Impact | Pressure (at which system fails) | n/a |
| Safety | [Not a mathematical concept in any scientific or engineering area we have identified]. Road safety is a product of collision probability and severity. Low values for either produce higher levels of safety on the roads (Sayed, Saunier, Lovegrove, & de Leur, 2010). | Unspecified threat origin, chronic | Impact | Probability of failing x consequence of failure (lower values are better) | (Over) Time |
| Security | The resources needed to overcome system protections. For instance, a 128 bit encryption key can be penetrated by a hacker with the resources to try $2^{128}$ passwords (Goldreich, 2005). | External, acute or chronic | Impact | Pressure (at which protection fails) | n/a |
| Sustainability | The continuous compensation of irreversible entropy production in an open system (Robinett, Rush, Wilson, & Reed, 2006). That is, the more sustainable a system, the less outside investment it requires to avoid degradation of functionality under normal operation. | Unspecified threat origin, chronic | Both/ neither | Investment (lower values are better) | (Over) Time |

# 5. PROPOSED DEFINITIONS AND VISUALIZATIONS

With the more concrete meanings obtained through the linguistic and mathematical framings of the concepts, we propose the following definitions for the nine concepts:

- Adaptability: Ability to change internally as necessary after a disruptive event to re-establish a high level of fitness.
- Agility: Ability to respond quickly to emerging challenges and opportunities.
- Reliability: Ability to perform within acceptable thresholds.
- Resilience: Capacity to recover critical functions and adapt following a disruptive event.
- Resistance: Capacity to absorb disruption with minimal damage to system functionality.
- Robustness: Capacity to withstand disruptions without damage to system functionality.
- Safety: Configuration that exposes the system to the fewest and least harmful disruptions.
- Security: Capacity to insulate the system from disruptions that would otherwise interrupt functionality.
- Sustainability: Ability to maintain a high level of functionality without inputs from external resources.

These system-specific definitions are visualized in the following nine charts (Fig. 1), each showing system performance over time according to greater or weaker manifestations of the different concepts. This visualization demonstrates that there are many different ways to achieve high system performance. Changes may be fast or far-reaching, and good performance may arise from observance of specific thresholds, an ability to withstand disruptions, or an ability to function without auxiliary resources.

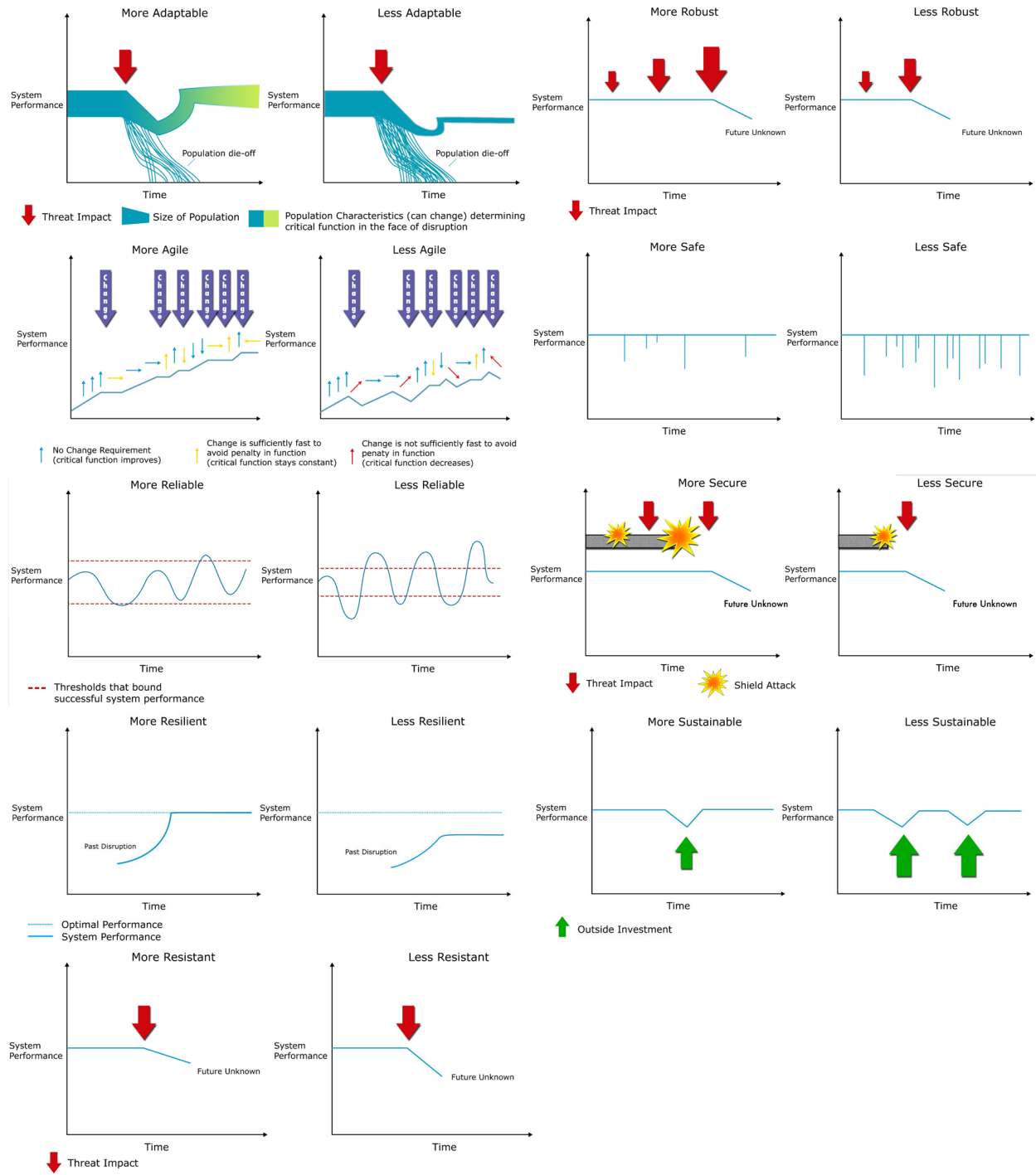

**Fig. 1.** Visualization of nine concepts considered in the study.

The concepts differ in their approaches to achieve good performance, but good performance itself is fairly consistent: relatively steady, and at a high level of performance. This demonstrates that the concepts are not mutually exclusive, indeed there are many opportunities to use them in concert. Below we examine each concept's relationship to the other concepts through its ability to contribute to their realization.

## 6. CONNECTING AND CONTRASTING THE CONCEPTS THAT CHARACTERIZE SYSTEMS WITHSTANDING THREATS

Using these results, we considered the relationships between the concepts, and how improving one can support another. The analysis below culminates with Figs. 2 and 3.

*Adaptability* can support internal reconfiguration that responds directly to the nature of disruptions faced, and high adaptability can thus support system improvements for realizing all the other concepts. It supports resilience using its ability to change to better accommodate new circumstances, allowing system recovery. Adaptability supports agility only if the change is fast, thus their relationship is conditional. Adaptability can increase system resistance, robustness, and security according to the changes needed to provide system protection. Over time, adaptability can allow the system to achieve a higher level of critical function in the new environment (safety). The ability to change incorporated in adaptability implies that internal reconfigurations will decrease the need for outside investment and thus improve overall sustainability and consistent functionality of the system (reliability).

*Agility*'s quick responses support adaptability because adaptability also involves change. High agility can support reliability, which prioritizes the stability that agility allows, and resilience, because high rates of needed change can spur recovery as well as avoid penalty. Agility is distinct from robustness or security, since both entail reliance on existing system protections rather than an ability to change to avoid damage. Agility can support resistance and safety by minimizing the consequences of a disruption, and can support sustainability by helping the system reorient in such a way to decrease the need for outside investment.

*Reliability* has no term about change so does not support adaptability or agility. It has time as a denominator, thus it can receive support from resilience, robustness, security, and resistance but does not support them. It could support safety if its thresholds align with safety priorities, but there are cases where reliability between suboptimal critical thresholds is more desirable or cost efficient than attempting to have high safety at all times, thus this relationship is conditional. Improved reliability also supports sustainability by limiting the declines of system functions and therein the outside investment needed to sustain it.

*Resilience* can only manifest when recovery is needed, and thus can complement concepts related to threat impact like resistance, robustness, safety, and security, but its improvement does not support them since recovery only existss after failure, which may be prevented by the aforementioned concepts. Improving resilience will improve reliability by minimizing time outside thresholds, and will improve sustainability by reducing the need for outside investment. Though adaptability and agility support resilience, resilience can support them only conditionally, because resilience may arise from either perseverance or change. Agility and adaptability capture a system's ability to recover by changing, but we note that there is no system concept that captures a system's ability to recover by perseverance alone. Perseverance can be implied in resilience, but is not synonymous with it.

Increased *resistance* improves reliability, safety, and sustainability by decreasing the amount the critical function declines when facing disruptions. Resistance does not involve aspects of change or consider the recovery process after a disruptions, thus it cannot support agility, adaptability and resilience. Resistance is only necessary if security is disabled, therefore its improvement would not affect security. Finally, resistance has a conditional relationship with robustness because they both represent an ability to withstand pressure, robustness in totality, and resistance by degrees.

*Robustness* prevents system functionality from experiencing disruptions, thus fully robust systems have no need for resilience, agility, or adaptability. Robustness can improve resistance by increasing the pressure necessary for a disruption to harm a system. Avoiding system harms will increase safety, reliability, and sustainability. Finally, robustness and security are similar, but the system components exhibiting them are different – robustness refers to the critical function itself, and security to some external protection that must first be compromised before the system itself is vulnerable. Therefore, improvements in robustness will not affect security.

*Safety* measures the probability and consequence of a disruption to a system's functionality, not the system's ability to withstand or mitigate the disruption (agility, resistance, robustness, security) or to recover from it (resilience, adaptability). Increasing safety means lowering the probability and consequence of a disruption, and supports system sustainability and reliability, thus safety supports those concepts.

*Security* can prevent disruptions from causing harm, thus supporting safety, reliability, sustainability. Improved security can negate the need for agility, adaptability, and resilience, thus improved security has no affect on these concepts. System security can support system robustness and resistance because it increases the overall pressure that can be withstood before a disruption harms the system.

Finally, *sustainability*, like reliability, is measured over time and is supported by many other concepts, but does not support them. Having more stable performance over time without dependence on outside contributions can increase measures of system safety and system reliability.

Fig. 2a below represents our evaluation for whether a concept column supports a concept row. The table is not symmetrical, and solid arrows denote relationships, while arrows that are not solid denote conditional relationships that cannot be definitively said to exist in all circumstances. These relationships are summarized in Fig. 2b, color-coded by whether concepts provide (blue) or receive (yellow) support from other concepts, or for the cases of reciprocal relationships (orange) where both concepts support each other, whether one of those relationships is conditional or not ("conditional reciprocity" in Fig. 2b).

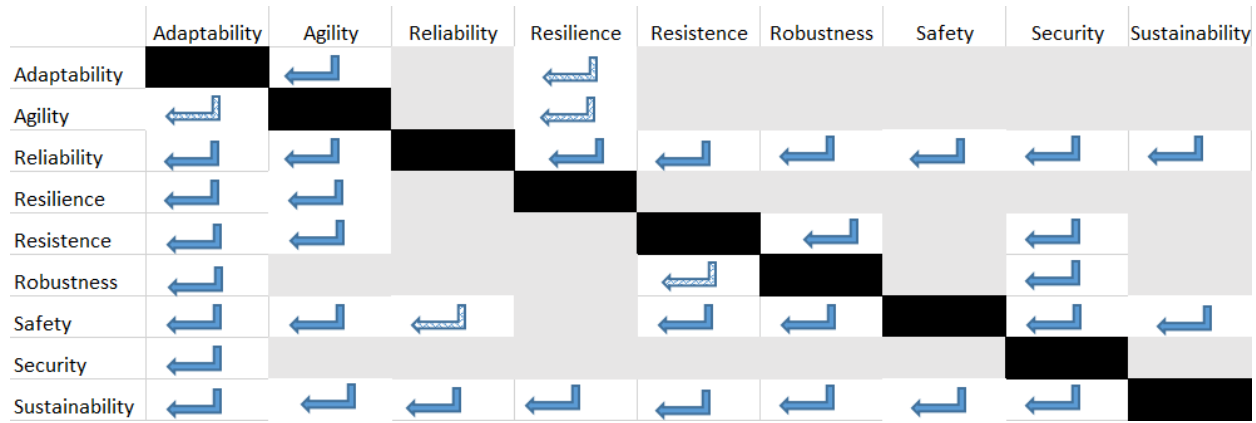

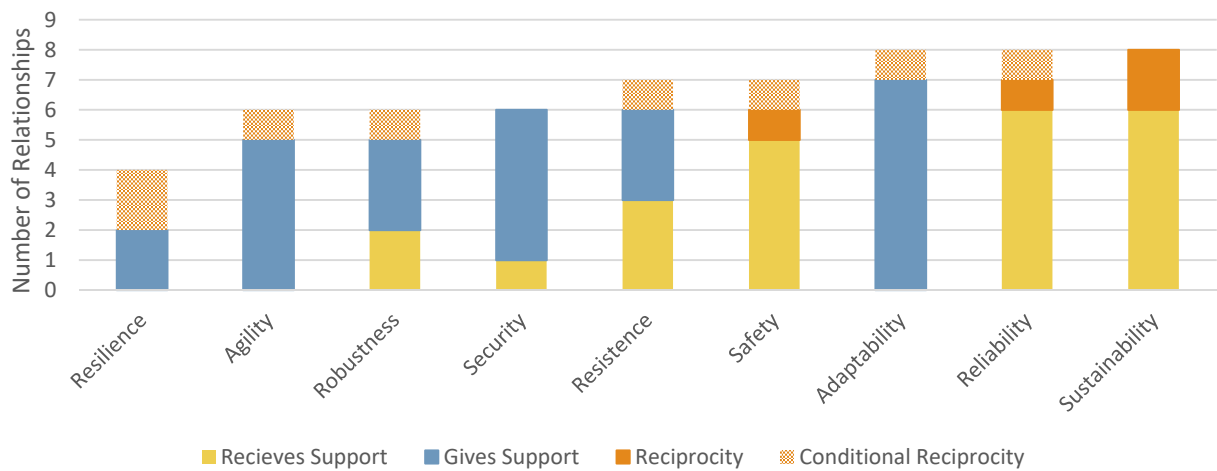

**Figs. 12a and 2b.** Summary of Relationships between Concepts. Fig. 2a (top) shows whether improving the ability of a system to realize one concept supports the system's ability to realize another concept. Fig. 2b (bottom) shows how often each concept provides or receives support in Fig. 2a, as well as reciprocal relationships.

We note that three concepts both give and receive support in a non-reciprocal manner (robustness, security, and resistance). The remaining six concepts (resilience, agility, safety, adaptability, reliability, and sustainability) all have reciprocal relationships (shown in orange) but otherwise either exclusively give support (blue) or exclusively receive support (yellow). All conditional relationships have reciprocity. We also note that the two concepts with the least relationships (resilience and agility) exclusively support others (with reciprocal relationships), whereas the three concepts with the most relationships are, beyond their reciprocal relationships, either exclusively providers of support (adaptability) or receivers of supports (reliability and sustainability). None of the extremes both give and receive non-reciprocal support.

There is correlation between the concepts that give support and the linguistic characterizations of concepts that denote action (Table II), with the exception of resilience. Similarly, there is a correlation between concepts receiving support and linguistic characterizations of overall performance, with the exception of safety, which we defined as a configuration, rather than as an ability, capacity, or measure of performance. The "reactions" of our linguistic analysis comprise two of the three concepts that both give and receive support in Fig. 2b. Though not perfectly correlated, Fig. 2b's distinctions between the concepts have much overlap with the distinctions made in Table II, and even more so with the more nuanced linguistic interpretation table in SI3. Therefore, we synthesize this information to propose a generalized conceptual framework relating to all the concepts (Fig. 3).

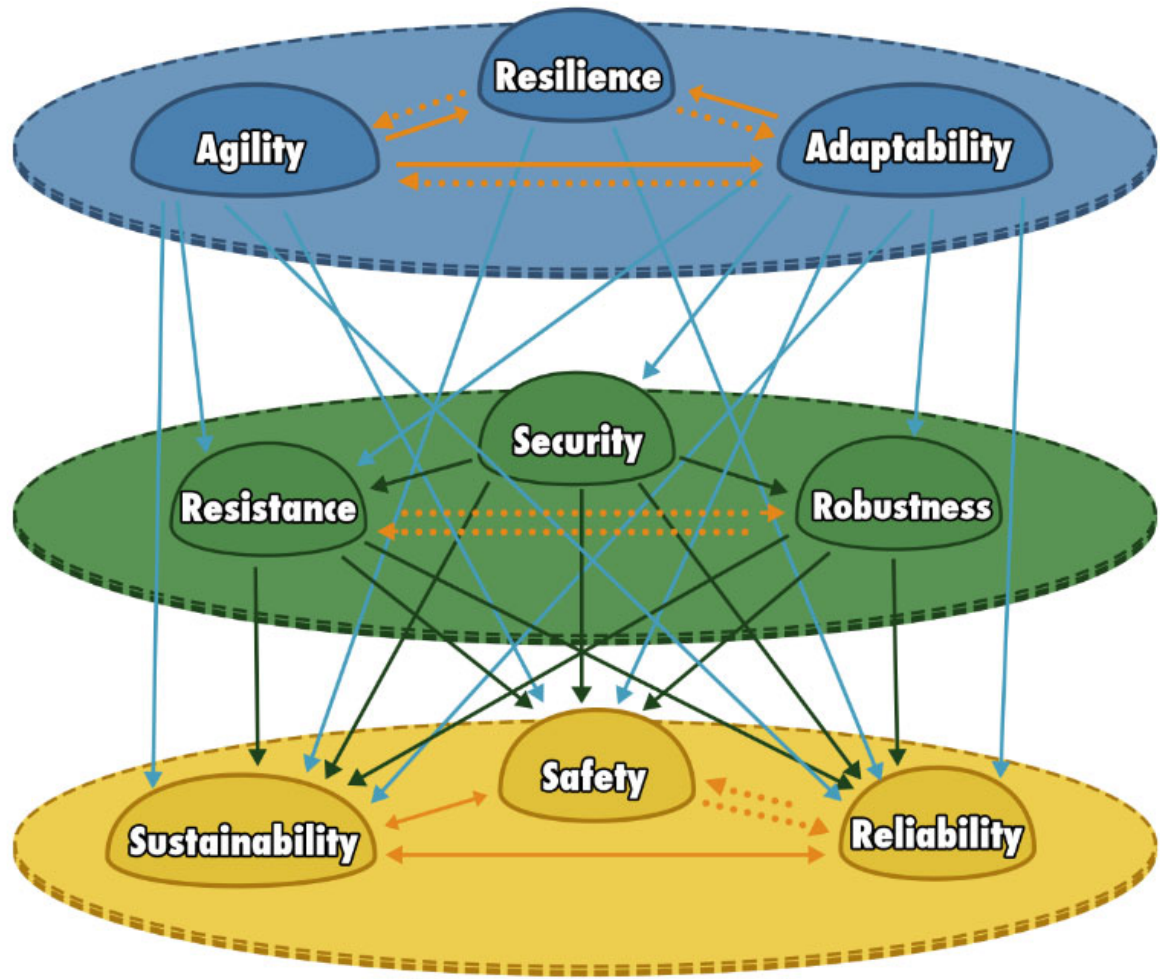

**Fig. 3.** Conceptual Framework Relating the Concepts

In Fig. 3, agility, adaptability, and resilience offer the most immediate reactions to an emerging threat, and in doing so provide support for the realization of the other concepts. Adaptability is especially noteworthy for the number of concepts it supports through its activation in a system. The action concepts are shown in the top tier, colored blue to indicate they provides support in accordance with Fig. 2b, with arrows to the concepts they support in lower tiers according to Fig. 2a. These supports are often mediated by the three concepts in the middle (security, robustness, resistance) that both give and receive support. Their tier is colored green to represent both providing and receiving support, colored blue and yellow respectively in Fig. 2b. We note that this layer corresponds to protective system infrastructure, which can be iteratively improved through actions taken in the top layer. Reliability, sustainability, and safety all receive support from the other concepts, and are colored yellow according to Fig. 2b. The reciprocal relationships, both conditional and unconditional (shown in orange), are always between concepts within the same level. The diagram generalizes the concepts according to levels that are more flexible than they appear, with some concepts acting as mediators in specific processes, but as inputs in others.

This analysis provides a structure for the process of attaining good system performance using these concepts: e.g., building the skill of agility can help a system become more adaptable, resilient, and reliable. We could view the top layer of supportive concepts as deliberate actions or choices taken, the middle concepts as short term consequences, abilities, or protections enabled by those actions, and at the bottom, the long-term outcomes. The outcomes, sustainability, reliability, and safety, all characterize the impacts of varying levels of performance over time, but not the reasons or processes that create them – these are provided by the concepts in the first two levels. This framework demonstrates that the concepts are not mutually exclusive: the structure shows where they support each other and even how they can be used in concert. The conceptual framework can reveal concrete pathways towards improving overall system performance, starting with the choices made at the moment the disruption emerges. Providing a better understanding and deliberately consistent usage of these concepts will enable systems managers to evaluate their systems, anticipate and validate the impacts of improvements, communicate their methodologies to other system managers using mutually comprehensible concepts.

## DISCLAIMER

The views, thoughts, and opinions expressed in the text belong solely to the authors, and not necessarily to the author's employer, organization, committee or other group or individual.